\documentclass[12pt]{article}
\usepackage{pdfpages}
\usepackage{afterpage}
 \usepackage{setspace}
\usepackage[ansinew]{inputenc}
\usepackage{graphicx}
\usepackage{color}
\usepackage{amssymb}
\usepackage{amsmath}

\newcommand{\cs}[3]{{{#3} \brace {#1 #2}}}
\newcommand{\h}[1]{\mathop{\lambda}\limits_{#1}\ \!\!\!}
\newcommand{\edf}{\ {\mathop{=}\limits^{\rm def.}}\ }

\newcommand{\al}{\alpha}
\newcommand{\s}[1]{\mathop{h}\limits_{#1}\ \!\!\!}

\begin{document}
{
\begin{center}{
{\large\bf{Motion of Charged Spinning Particles in a Unified Field}}
}\end{center}
\begin{center}
  M. I. Wanas$^{*,}$\footnote{Egyptian Relativity Group (ERG), Cairo, Egypt.} and Mona M. Kamal$^{**,1}$
\end{center}
\begin{center}
  $^*$ Astronomy Department, Faculty of Science, Cairo University, Egypt.\\
  E-mail:wanas@scu.eg
\end{center}
\begin{center}
  $^{**}$ Mathematics Department, Faculty of Girls, Ain Shams University, Egypt.\\
  Email:elhoseinymona@gmail.com
\end{center}
\begin{center}
{\bf Abstract}
\end{center}
Using a geometry wider than Riemannian one, the  parameterized absolute parallelism (PAP-) geometry, we derived a new curve
containing two parameters. In the context of the geometrization philosophy, this new curve can be used as a trajectory
of charged spinning test particle in any unified field theory constructed in the PAP-space.
We show that imposing certain conditions on the two parameters, the new curve can be reduced to
a geodesic curve giving the motion of a scalar test particle or/and a modified geodesic giving the motion
  of neutral spinning test particle in gravitational field. The new method used for derivation, the Bazanki
  method, shows a new feature in the new curve equation.  This feature is that the equation contains the
  electromagnetic potential term together with the Lorentz term. We show the importance
  of this feature in physical applications. 
\section{Introduction}
According to the geometrization philosophy the curve in a certain geometry represents the equation of motion of a
theory which constructed in this geometry.
Together with the field equations of any theory we need the equation of motion which characterize the theory used.
In general relativity, geodesic curve is considered as an equation of motion of a scalar test particle moving in a gravitational field.

Geodesic equation can be derived using the Lagrangian (cf. \cite{1})
\begin{equation}\label{1}
 L_1\edf ~g_{\mu\nu}~\dot{x}^\mu~\dot{x}^\nu,\end{equation}
 where $g_{\mu\nu}$ is the metric tensor and $\dot{x}^\mu(\edf \frac{dx^\mu}{ds})$ is the unit tangent vector to the curve.
Euler-Lagrange equation is given by (cf. \cite{2})
\begin{equation}\label{2}
 \frac{d}{ds}~\frac{\partial L_1}{\partial \dot{x}^\gamma}-\frac{\partial L_1}{\partial x^\gamma}=0,\end{equation}
such that $s$ is the scalar parameter varying along the curve. Using Lagrangian (\ref{1}) and equation (\ref{2}) we get
\begin{equation}\label{3}
\ddot{x}^\alpha+ \cs{\mu}{\nu}{\alpha}\dot{x}^\mu \dot{x}^\nu=0,\end{equation}
where $\cs{\mu}{\nu}{\alpha}$ is the coefficient of Levi-Civita  linear connection which is defined as
\begin{equation}\label{cs}
\cs{\mu}{\nu}{\alpha} \edf \frac{1}{2} g^{\alpha \sigma} (g_{\mu\sigma,\nu}+g_{\nu\sigma,\mu}-g_{\mu\nu,\sigma}).
\end{equation}
Equation (\ref{3}) is the curve equation of the Riemannian geometry.

The Lagrangian used for deriving the equation of motion of a charged particle moving in the presence of electromagnetic field is defined by (cf. \cite{3})
 \begin{equation}\label{4}
   L_2 \edf g_{\mu\nu}(V^\mu + \beta A^\mu)V^\nu,
 \end{equation}
where $A_\mu$ is a vector field and $\beta$ is a conversion parameter given by
  \begin{equation}\label{beta}
    \beta\edf\frac{e}{m},
  \end{equation}
  where $e$ is the electron charge and $m$ is the
electron  mass.
  Using Lagrangian (\ref{4}) and Euler-Lagrange equation (\ref{2}),
we get the equation of motion of a charged test particle moving  in a combined gravitational and electromagnetical fields (cf. \cite{s84})
 \begin{equation}\label{5}
   \frac{dV^\alpha}{ds} +\cs{\mu}{\nu}{\alpha} V^\mu V^\nu = - \beta F^\alpha_{.~\nu}V^\nu ,
 \end{equation}
 where $F_{\mu\nu}$ is the curl of the vector $A_\mu$. The term on the right hand side of equation (\ref{5}) is known as a Lorentz force term.


In 1989, Bazanski \cite{B89} suggested one Lagrangian to derive both geodesic and geodesic deviation equations in Riemannian geometry, which is given by
\begin{equation}\label{lb}
 L_B\edf g_{\mu\nu}~V^\mu~\frac{D\Psi^\nu}{Ds},\end{equation}
where $V^\mu$ is the unit vector tangent to the path, $\Psi^\nu$ is the deviation vector and
 \begin{equation}\label{o}
\frac{D\Psi^\nu}{Ds}\edf \Psi^\nu_{.;\alpha}V^\alpha.\end{equation}
 The semicolon operator $(;)$ denotes covariant differentiation using Levi-Civita connection while the comma $(,)$ stands for ordinary partial differentiation. According to Bazanski,  variation with respect to deviation vector $\Psi^\mu$ gives geodesic equation, while variation with respect to the unit vector $V^\mu$ gives geodesic deviation equation.

   Riemannian geometry has a unique linear connection, which is Livi-Civita connection. Wanas et. al. \cite{We95} have modified  Bazanski approach  in a different geometry. This geometry  has other linear connections together with the Livi-Civita connection.
   They have applied Bazanski Lagrangian using the four linear connections defined in the absolute parallelism (AP-) geometry.
    The four connections are: Weitzenb$\ddot{o}$ck  connection $\Gamma^\alpha_{~\mu\nu}$, dual connection  $\tilde{\Gamma}^\alpha_{~\mu\nu}(\edf \Gamma^\alpha_{~\nu\mu} )$,  $\Gamma^\alpha_{~(\mu\nu)}(\edf \frac{1}{2}(\Gamma^\alpha_{~\mu\nu}+\Gamma^\alpha_{~\nu\mu} ))$ and Livi-Civita connection.
    Using each connection, they have got  new definitions for operator $\frac{D}{Ds}$ which appears in the Bazanski Lagrangian (\ref{lb}).
  Wanas et. al.  have obtained a new set of three different path equations.
   The three different path equations can be written as \cite{We95}
   \begin{equation}\label{e9}
  \frac{dJ^\mu}{ds^-}+ \cs{\alpha}{\beta}{\mu}J^\alpha J^\beta=0,
\end{equation}
\begin{equation}\label{e10}
  \frac{dW^\mu}{ds^0}+ \cs{\alpha}{\beta}{\mu}W^\alpha W^\beta=-~\frac{1}{2}~{\Lambda}^{~~~~\mu}_{(\alpha\beta).}~W^\alpha W^\beta,
\end{equation}
\begin{equation}\label{e11}
  \frac{dV^\mu}{ds^+}+ \cs{\alpha}{\beta}{\mu}V^\alpha V^\beta=-~{\Lambda}^{~~~~\mu}_{(\alpha\beta).}~V^\alpha V^\beta,
\end{equation}
 where $V^\mu$, $W^\mu$ and $J^\mu$ are unit tangent vectors to the curves characterized by parameters
 $s^+$, $s^0$ and $s^-$, respectively.

If the moving particle has another property, for example, like spin, then geodesic equation (\ref{3}) isn't suitable for describing the motion of such  particle.
An important property of the set of equations (\ref{e9}), (\ref{e10}) and (\ref{e11}) appears in its right hand side. This property is the jumping parameter of the right hand side of the above mentioned equations. This property motivates Wanas  \cite{w98} to consider the right hand side of the above set of equations as representing a geometric interaction between the quantum spin of the moving particle and  the torsion of the background geometry. For this property, Wanas generalized the AP-space by constructing a new version called the parameterized absolute parallelism (PAP).

Parameterized absolute parallelism (PAP-) geometry \cite{w98} has spectrum  of spaces. It can be reduced to Riemannian and absolute parallelism geometries in some special cases. Applying the modified Bazanski approach in the context of PAP-geometry, Wanas \cite{w98} obtained a modified geodesic equation.
$$\frac{dV^\mu}{ds}+ \cs{\alpha}{\beta}{\mu}V^\alpha V^\beta =-~b \Lambda^{~~~~\mu}_{(\alpha\beta).}~V^\alpha V^\beta.$$
This equation describes the motion of a spinning particle moving in a gravitational field. This equation can reduce to geodesic one in a special case ($b=0$).

In the framework of PAP-geometry we are going to derive the equation of  motion of a spinning and charged particle moving in a combined gravitational and electromagnetic fields (a unified field), using modified Bazanski approach.

\section{Geometry Used: Parameterized Absolute Parallelism Geometry}

This work is carried out in the context of the "\textit{Parameterized Absolute Parallelism}" (PAP-) geometry  abbreviated as $(M, \h{i})$ \cite{w98}. M is  an n-dimensional differentiable  manifold and $\h{i}$ is a set of  n-independent  vector fields.  The components of these vector fields are   considered as the building blocks (BB)\footnote{BB are geometric quantities using which we can construct all objects of the geometry.} of  this geometry as it is considered in the AP-geometry.  Since the determinant  $\h{}~(=|\h{i}_{\mu}|)$ is non-vanishing, i.e. $\h{}\neq0$, then the covariant components  $\h{i}_\mu(x)$ \footnote{We use Greek indices for coordinate components written in a covariant or contravariant positions. Latin indices are used to represent vector numbers, written always in a lower position. Summation convention is carried out for Greek indices in the usual way,
 while for Latin dummy indices the operation is carried out wherever the indices
   appear in the same term.} and contravariant $\h{i}^\mu$ components satisfy the following relations
  \begin{equation}\label{}
\h{i}^\alpha \h{i}_\beta = \delta^{\alpha}_{~\beta},\end{equation}
and
\begin{equation}\label{}
 \h{i}^\alpha \h{j}_\alpha = \delta_{ij},\end{equation}
where $\delta_{ij}$ is the Kronecker delta.
 We can define, from which, the following second order symmetric tensors,
\begin{equation}\label{gl}
g_{\mu \nu}  \edf \h{i}_{\mu}\h{i}_{\nu},\end{equation}
and
\begin{equation}\label{gu}
g^{\mu \nu}  \edf \h{i}^{\mu}\h{i}^{\nu}.\end{equation}
Consequently,
$$g^{\alpha\mu}g_{\alpha\nu}=\delta^\mu_{~\nu}. $$

The second order tensor $g_{\mu\nu}$ can be used as the metric
 tensor to define, as a special case,  a Riemannian space in the  context of the PAP-geometry.

 The PAP- linear connection is given by \cite{w98},
 \begin{equation}\label{pconn}
\nabla^{\al}_{.~\mu\nu}\edf\cs{\mu}{\nu}{\alpha}~+{\mathop{\gamma}\limits^{ {{*}}}}~^{\alpha}_{.~\mu\nu}~,\end{equation}
where  ${\mathop{\gamma}\limits^{ {{*}}}}~^{\alpha}_{.~\mu\nu}$ is a third order tensor, called the parameterized contortion, defined by
\begin{equation}\label{PHS}
{\mathop{\gamma}\limits^{ {{*}}}}~^{\alpha}_{.~\mu\nu}\edf b ~\gamma^{\alpha}_{.~\mu\nu}= b  \h{i}^{\alpha} \h{i}_{\mu;\nu},
\end{equation}
such that $b$ is a dimensionless parameter.

An important note is that inserting the parameter $b$ will not cause any difference to the
properties of the AP-space, but will add to them, for example the Riemannian geometry is
defined in the AP-space as an associated space to it, but in the present case it is can be
considered as a special case. This will be more clear in the text.

The parameterized connection (\ref{pconn}) has been proved to be a metric one \cite{w98}. i.e. satisfies metricity condition\footnote{We use the double stroke $_{||}$ and (+) sign to characterize covariant differentiation using the parameterized connection (\ref{pconn}).}
\begin{equation}\label{mc}
 g_{\stackrel{\mu\nu||\sigma}{++~~~}}\equiv0. \end{equation}
Since  $ \nabla^{\alpha}_{.~\mu\nu}$ is non-symmetric, then the parameterized torsion tensor $ {\mathop{\Lambda}\limits^{ {{*}}}}~^{\alpha}_{.~\mu\nu}$ is  defined by,
\begin{equation}\label{cl}
\begin{array}{cl}
{\mathop{\Lambda}\limits^{ {{*}}}}~^{\alpha}_{.~\mu\nu}&\edf \nabla^{\alpha}_{.~\mu\nu}-\nabla^{\alpha}_{.~\nu\mu}  \\
&={\mathop{\gamma}\limits^{ {{*}}}}~^{\alpha}_{.~\mu\nu}-{\mathop{\gamma}\limits^{ {{*}}}}~^{\alpha}_{.~\nu\mu} =b \Lambda^{\al}_{.~\mu\nu},
\end{array} \end{equation}
 and
  $$\Lambda^{\al}_{.~\mu\nu}\edf \gamma^{\alpha}_{.~\mu\nu}-\gamma^{\alpha}_{.~\nu\mu},  $$
 is the torsion tensor of the AP-space.
 The contraction of the parameterized torsion (\ref{cl}) or contortion, is given by,
\begin{equation}\label{ccl}
{\mathop{c}\limits^{ {{*}}}}_\mu\edf{\mathop{\Lambda}\limits^{ {{*}}}}~^{\alpha}_{.~\mu\alpha}
={\mathop{\gamma}\limits^{ {{*}}}}~^{\alpha}_{.~\mu\alpha}=b~c_\mu.
\end{equation}

Also, due to the non-symmetry of the parameterized connection (\ref{pconn}), there exist two more linear connections: The dual connection  defined by
\begin{equation}\label{}
 \tilde{\nabla}^{\al}_{.~\mu\nu}\edf\nabla^{\al}_{.~\nu\mu},  \end{equation}
  and the symmetric part of $\nabla^{\al}_{.~\mu\nu}$  which is given as
\begin{equation}\label{pcsym}
\begin{array}{cl}
\nabla^{\alpha}_{.~(\mu\nu)}&\edf\frac{1}{2} (\nabla^{\alpha}_{.~\mu\nu}+\nabla^{\alpha}_{.~\nu\mu}) \\\\
&~=\cs{\mu}{\nu}{\alpha}+~\frac{1}{2}~{\mathop{\Delta}\limits^{ {{*}}}}~^{\alpha}_{.~\mu\nu},\end{array}
\end{equation}
where
 $${\mathop{\Delta}\limits^{ {{*}}}}~^{\alpha}_{.~\mu\nu}\edf {\mathop{\gamma}\limits^{ {{*}}}}~^{\alpha}_{.~\mu\nu}+{\mathop{\gamma}\limits^{ {{*}}}}~^{\alpha}_{.~\nu\mu}.$$

The PAP-geometry has four linear connections which are $\nabla^{\al}_{.~\mu\nu}$, $\tilde{\nabla}^{\al}_{.~\mu\nu}$,  $\nabla^{\al}_{.~(\mu\nu)}$ and $\cs{\mu}{\nu}{\alpha}$. So, we have four different curvature tensors, defined by ordinary manner, using the commutation elation for each one of these connections.

For each connection we can define the following tensor derivatives
\begin{equation}\label{22}
K^{\stackrel {\alpha}{+}}_{~.~||\beta}\edf K{^\alpha}_{,\beta}+K^{\mu} \nabla{^\alpha}_{\mu\beta},
\end{equation}
\begin{equation}\label{nd}
K^{\stackrel {\alpha}{-}}_{~.~||\beta}\edf K{^\alpha}_{,\beta}+K^{\mu} \nabla{^\alpha}_{\beta\mu},
\end{equation}
\begin{equation}
K^ {\alpha}_{~.~||\beta}\edf K{^\alpha}_{,\beta}+K^{\mu} \nabla{^\alpha}_{(\mu\beta)},
\end{equation}
where $K^\alpha$ is any arbitrary vector field defined in the PAP-space.

The curve equation characterizing PAP-geometry is given by, as mentioned above is \cite{w98}
\begin{equation}\label{modgd}
  \frac{dV^\mu}{d\tau}+ \cs{\alpha}{\beta}{\mu}V^\alpha V^\beta=-~b~{ \Lambda}^{~~~~\mu}_{(\alpha\beta).}~V^\alpha V^\beta,
\end{equation}
where
\begin{equation}\label{}
  { \Lambda}^{~~~~\mu}_{(\alpha\beta).}\edf \frac{1}{2}({ \Lambda}^{~~~\mu}_{\alpha\beta.}+{ \Lambda}^{~~~\mu}_{\beta\alpha.}).
\end{equation}
The PAP-geometry reduces to the Riemannian one if we take $b=0$, while, for $b=1$ the PAP-geometry reduces to AP-geometry.
 At any stage of calculations we can go back to Riemannian or AP geometries as two special cases.
 The dimensionless  parameter $b$ is suggested, for physical applications,  to take the value \cite{w98}
\begin{equation}\label{b}
b\edf\frac{N}{2}\alpha \gamma,
\end{equation}
where $N$ is an integer number takes the values $(N=0,~ 1,~ 2,~ ...)$, $\alpha$ is the fine structure constant and $\gamma$ is a dimensionless parameter to be adjusted with experiments or observations for every system.

\section{Path Equation for Charged and Spinning Particles}
In  order to derive a general equation of motion let us define the parameterized Lagrangian  in the PAP-geometry by

\begin{equation}\label{2.lm}
 L\edf g_{\mu\nu}~(U^\mu+\beta {c}^\mu)~\frac{D\psi^\nu}{D\tau}~,\end{equation}
where $U^\mu$ is the tangent to a path characterized by the parameter $\tau$,
$\psi^\nu $ is the deviation vector and
$$ \frac{D\psi^\nu}{D\tau}\edf\psi^{{\nu}}_{\stackrel{.~||\alpha}{+~~~~~}} ~U^\alpha ~,$$
so, the Lagrangian (\ref{2.lm}), can be written in the explicit  form
\begin{equation}\label{25}
 L = g_{\mu\nu}~(U^\mu+\beta {c}^\mu)~U^\alpha(\psi^\nu_{.~,\alpha}~+~\psi^\epsilon ~
 \nabla^\nu_{.~\epsilon\alpha})~.\end{equation}
Now it is clear that the Lagrangian (\ref{25}) has two parameters: One is the dimensionless parameter $b$ mentioned above
 in the parameterized connection $ \nabla^\nu_{.~\epsilon\alpha}$ and the other is $\beta$.

By varying the Lagrangian (\ref{2.lm}) with respect the  deviation vector $\psi^\gamma$, we have
\begin{equation}\label{2.t1}
\frac{\partial L }{\partial \psi^\gamma}~=~ g_{\mu\nu}~(U^\mu+\beta {c}^\mu)~U^\alpha~\nabla^\nu_{.~\gamma\alpha}~, \end{equation}
\begin{equation}\label{}
\frac{\partial L }{\partial \dot{\psi}^\gamma}~=~ g_{\mu\gamma}~(U^\mu+\beta {c}^\mu), \end{equation}
and
\begin{equation}\label{2.t2} \frac{d}{d\tau}~\frac{\partial L }{\partial \dot{\psi}^{ \gamma}}~
=~g_{\mu\gamma}~\frac{d U^\mu}{d\tau}~+~g_{\mu\gamma,\sigma}~U^\mu~U^\sigma
+\beta \frac{d {c}_\gamma}{d\tau}~.\end{equation}
Substituting from (\ref{2.t1}) and (\ref{2.t2}) into the Euler-Lagrange equation
$$ \frac{d}{d\tau}~\frac{\partial L }{\partial \dot{\psi}^{ \gamma}}-\frac{\partial  L}{\partial \psi^\gamma}=0,$$
we get,
$$\label{}
g_{\mu\gamma}~\frac{d U^\mu}{d\tau}~+~g_{\mu\gamma,\sigma}~U^\mu~U^\sigma
+\beta \frac{d {c}_\gamma}{d\tau}~- g_{\mu\nu}~(U^\mu+\beta {c}^\mu)~U^\alpha~\nabla^\nu_{.~\gamma\alpha}=0,
$$
from metricity condition (\ref{mc}), we obtain
\begin{equation}\label{32}
\dot{U}^\mu+\nabla^\mu_{.~\gamma\alpha}U^\alpha~U^\gamma
~=~-\beta \dot{c}^\mu+~\beta {c}_\nu~\nabla^\nu_{.~\gamma\alpha}U^\alpha~g^{\gamma\mu}.
\end{equation}
Using definitions (\ref{pconn}) and (\ref{22}),  equation (\ref{32}) can be written, explicitly, in the form,
\begin{equation}\label{2.mpath}
\frac{dU^\mu}{d\tau}+ \cs{\alpha}{\beta}{\mu}U^\alpha U ^\beta=-~b~{\gamma}^{\mu}_{.(\alpha\beta)}~U^\alpha U ^\beta
-\beta ~g^{\nu\mu}~c_{\stackrel{\nu||\alpha}{+~~~}}U^\alpha.
\end{equation}
It is obvious that if $b=0$ and $\beta=0$ the modified  equation (\ref{2.mpath}) reduces to the geodesic equation (\ref{3}). Also, if $b=1$ these equations tends to the equations (\ref{5}) if we use the conventional method. When the electromagnetic sector  is switched off   (i.e. $\beta=0$) the parameterized path equations (\ref{2.mpath}) reduces to (\ref{modgd}).
 We can take the parameter $\beta=\frac{e}{m}$ where $e$ is the electron charge and $m$ is the electron mass. This is done for dimensional consideration of the Lagrangian (\ref{2.lm}). Note that $ c^\mu$ is considered as a geometric representation of the electromagnetic potential.

We can rewrite equation (\ref{2.mpath}) as follows
\begin{equation}\label{new}
\frac{dU^\mu}{d\tau}+ \cs{\alpha}{\beta}{\mu}U^\alpha U ^\beta=-~b~{\gamma}^{\mu}_{.(\alpha\beta)}~U^\alpha U ^\beta
-\beta ~g^{\nu\mu}~U^\alpha~F_{\nu\alpha}-\beta ~g^{\nu\mu}~U^\alpha~c_{\stackrel{\alpha||\nu}{-~~~}},
\end{equation}
 where
 \begin{equation}\label{fd}
   F_{\nu\alpha}\edf c_{\nu,\al}-c_{\al,\nu}.
 \end{equation}
Important note: It is to be considered that equation (\ref{fd}) is a field equation
to be solved and a definition after the solution. This point will be more discussed in
Section 5.
\section{Linearization of The New Path Equation}
In the context of geometrization philosophy, the BB of the geometry are considered as the field variables of the theory. So,
in order to gain more physical meaning from the two parameter geometric equation (\ref{new}), we are going first to linearize it.
 We consider  a weak and a static unified field together with a slowly moving particle in this field, i.e. we assume
\begin{enumerate}
  \item  The BB ($\h{i}_\mu$) of the PAP-geometry in the form
\begin{equation}\label{w}
\h{i}_\mu=\delta_{i\mu}+\epsilon \s{i}_\mu,\end{equation}
where  $\s{i}_\mu$ represents a set of functions of the coordinates causing deviation from Euclidean geometry.
We expand all geometric quantities needed to the first order in $\epsilon$ and neglecting higher orders.
  \item A Static unified field, which means that
  \begin{equation}\label{}
   \frac{\partial \h{i}_\mu}{\partial x^0}=0.
  \end{equation}
  \item A slowly moving particle, i.e. $U^0\simeq 1,~U^1\sim U^2\sim U^3\sim \varepsilon$.
Also, $O(\varepsilon^2)$, $O(\varepsilon \epsilon)$ and higher order can be neglected.
\end{enumerate}
As a consequence of (\ref{w}), we get
\begin{equation}\label{wi}
\h{i}^\mu=\delta_{i\mu}-\epsilon \s{i}^\mu .\end{equation}
  Substituting from (\ref{w}) and (\ref{wi}) into (\ref{gl}) and (\ref{gu}), we obtain
  $$g_{\mu\nu}=\delta_{\mu\nu}+\epsilon~y_{\mu\nu},$$
   $$g^{\mu\nu}=\delta_{\mu\nu}-\epsilon~y_{\mu\nu},$$
  and \begin{equation}\label{}
     y_{\mu\nu} \edf(\s{\mu}_\nu+\s{\nu}_\mu).
      \end{equation}

The linearized symmetric part of the  contortion tensor $\gamma ^\mu_{.~\alpha \beta}$  will be in the form
 \begin{equation}\label{}
 {\gamma}^{\mu}_{.~(\alpha\beta)}=\frac{\epsilon}{2}~(y_{\alpha\beta,\mu}-\s{\alpha}_{\mu,\beta}-\s{\beta}_{\mu,\alpha}),
  \end{equation}
  also, the first order of the contracted  torsion or contortion 
  becomes
  \begin{equation}\label{wc}
 c_{\mu}=\epsilon~(\s{\al}_{\mu,\al}-\s{\al}_{\al,\mu}).
  \end{equation}
  Recalling the definition (\ref{nd}) and (\ref{fd}), equation (\ref{new}) becomes
\begin{equation}\label{nl}
\frac{dU^\mu}{d\tau}+ \cs{\alpha}{\beta}{\mu}U^\alpha U ^\beta=-~b~{\gamma}^{\mu}_{.(\alpha\beta)}~U^\alpha U ^\beta
-\beta ~g^{\nu\mu}~(c_{\nu,\alpha}-c_\sigma~\cs{\alpha}{\nu}{\sigma}- b~c_\sigma ~\gamma ^\sigma_{~\nu\alpha})U^\alpha.
\end{equation}
Applying the assumptions 1, 2 and 3 to the equation of motion (\ref{nl}), we get
\begin{equation}\label{h}
\frac{dU^\mu}{dt}~=~-~ \cs{0}{0}{\mu}-~b~{\gamma}^{\mu}_{.~00},
\end{equation}
 $$ ~~~~~~~~~~~~~~=\frac{\epsilon}{2}~y_{00,\mu}-b~\frac{\epsilon}{2}~y_{00,\mu}, $$
which can be written in terms of Newtonian potential as
    \begin{equation}\label{l6}
   \frac{d^2x^a}{dt^2}=(1-b)\Phi_{,a}~~~  a=1,~2,~3.
 \end{equation}
where $$
   \Phi\edf\frac{\epsilon}{2}~y_{00},$$
 is the Newtonian gravitational potential.
\section{Discussion and Concluding Remarks}
In the framework of a unified field theories, we expect to produce gravity from electromagnetism and
vice versa. The first production is well known since Einstein-Maxwell theory \cite{1}. Nature provides
us with some evidences for producing electromagnetism from gravity, since  most (if not all) celestial objects
have magnetic fields of different orders while they are all electrically neutral. Some authors \cite{me95} have
found a theoretical relation between the magnetic field and some gravitational properties. Other authors
\cite{dse10} have used this relation to interpret the huge magnetic field producing gamma ray bursts.

Now, in the present work, we are dealing with a pure geometric theory unifying gravity  and electromagnetism \cite{WK14}.
 So, we expect both types\footnote{As will appear in this discussion two types of electromagnetism appear in the Universe.} of  electromagnetism to be present in the theory this will be clear in applications.
 In what follows we call the first type {\bf cosmic electromagnetic field} and the second type {\bf conventional electromagnetic field}.
 Both will affect the motion of particles moving in the unified field. As usual procedure we solve the field equations before solving
 the equations of motion as usually done in GR. In the present work the field equations \cite{WK14} are in general sixteen in
 number while the equations of motion (\ref{new}) are four.  After solving the field equation, sixteen field variables become known
 function of the coordinate. Afterwards we solve the equations of motion (\ref{new}) to get the components of the acceleration (or velocity)
 of the moving particles.

In the context of the geometrization philosophy, any field theory contains
two types of equations. The first type controls the behaviour of the field
 (field equations) originated from the differential identities of the geometry
 used. The second type governs the motion of a particle in the field mentioned
 above (equations of motion). For example, in general relativity the field
 equations emerged from Bianchi differential identity of Rimannian geometry, while the equations of motion are the general curves of the same geometry (geodesic).

 In the present work, we use another theory \cite{WK14} written in the PAP-geometry \cite{w98}.
 Here, we derive the equation of the general curve in the PAP-geometry (\ref{new}) which is used as an equation of motion of the theory. The field theory completed here has curvature and anti-curvature. It has been shown that curvature gives rise to gravity and anti-curvature gives rise to anti-gravity \cite{W12}. These theoretical predictions of equation (\ref{modgd}), in its linearized form, have been supported by interpreting the discrepancy of the COW-experiment \cite{c75} \cite{c80}\cite{c88}, which is verified in 2000.

Equation (\ref{new}) is the general equation of motion for an electrically charged and spinning test particle moving in a general field unifying gravity and electromagnetism  \cite{WK14}. This equation contains  two parameters $b$ and $\beta$. The first parameter is related to the quantum spin of the moving particle and the second is related to the electric charge of the moving particle. Equation (\ref{new}) has the following properties.
\begin{enumerate}
  \item If the two parameters $b$ (\ref{b}) and $\beta$ (\ref{beta}) simultaneously vanish identically, i.e. the moving particle is a scalar (electrically neutral and has zero quantum spin). In this case (\ref{new}) reduces to an ordinary geodesic (\ref{3}), of Riemannian geometry.
  \item If the parameter $\beta$ (or $e$) vanishes alone, in this case equation (\ref{new}) reduces to equation (\ref{modgd}) for a spinning particle.
  \item If the parameter $b=0$, then (\ref{new}) will reduce to 
\begin{equation}\label{z}
\frac{dU^\mu}{d\tau}+ \cs{\alpha}{\beta}{\mu}U^\alpha U ^\beta=
-\beta ~g^{\nu\mu}~U^\alpha~F_{\nu\alpha}-\beta ~g^{\nu\mu}~U^\alpha~c_{{\alpha;\nu}},
\end{equation}
which is the equation of motion for a charged particle derived using the Bazanski scheme.
\end{enumerate}
In the present work, we applied general scheme called linearization of a non-linear field theory. This scheme is applied after generalization to get the forces affecting the motion of an electrically charged and spinning test particle in a field unifying gravity and electromagnetism.

The conventional electromagnetic field can be present together with cosmic electromagnetic field in the
context of some terrestrial experiments. This result gives the geometric interpretation of the
Aharonov-Bohm effect. To get same result from the present work, more efforts are needed.

\end{document}